\documentclass[altaffilletter,aps,nofootinbib,twocolumn,prd,eqsecnum,preprintnumbers,superscriptaddress]{revtex4-1}
\pdfoutput=1
\usepackage[caption=false]{subfig}
\usepackage{graphicx}
\usepackage{amsmath}
\usepackage{amsfonts}
\usepackage{amssymb}
\usepackage{color}
\usepackage{bm}
\usepackage{mathrsfs}
\usepackage{epstopdf}
\usepackage{url}
\usepackage{footnote}
\usepackage{textcomp}
\usepackage{dsfont}
\usepackage{ulem}
\usepackage{hyperref}
\usepackage{enumerate}

\makeatletter
\newcommand*{\rom}[1]{\expandafter\@slowromancap\romannumeral #1@}
\makeatother

\begin{document}

\title{Eikonal black hole ringings in generalized energy-momentum squared gravity}

\author{Che-Yu Chen}
\email{b97202056@gmail.com}
\affiliation{Department of Physics and Center for Theoretical Sciences, National Taiwan University, Taipei, Taiwan 10617}
\affiliation{LeCosPA, National Taiwan University, Taipei, Taiwan 10617}

\author{Pisin Chen}
\email{pisinchen@phys.ntu.edu.tw}
\affiliation{Department of Physics and Center for Theoretical Sciences, National Taiwan University, Taipei, Taiwan 10617}
\affiliation{LeCosPA, National Taiwan University, Taipei, Taiwan 10617}
\affiliation{Kavli Institute for Particle Astrophysics and Cosmology, SLAC National Accelerator Laboratory, Stanford University, Stanford, California 94305, USA}
\begin{abstract}
In the scope of black hole spectroscopy, several attempts have been made in the past decades to test black holes or gravitational theories via black hole quasinormal modes. In the eikonal approximation, the quasinormal modes are generically associated with the photon ring of the black hole. This correspondence is valid for most cases in general relativity, but may not be true in other theories of gravity. In this paper, we consider the generalized energy-momentum squared gravity in which matter fields are nonminimally coupled to geometry. We investigate the axial perturbations of the charged black holes in this model, without assuming any explicit expression of the action functional. After obtaining the modified Klein-Gordon equation and the modified Maxwell equations, we perturb the gravitational equations and the modified Maxwell equations to derive the coupled master equations of the axial perturbations. In the presence of the nonminimal coupling between matter and geometry, the correspondence between the eikonal quasinormal modes and the photon ring is not satisfied in general. Also, the two coupled fields of the axial perturbations are found to propagate independently, and they do not share the same quasinormal frequencies in the eikonal limit. 
\end{abstract}

\maketitle

\section{Introduction}   

Several observational pieces of evidence have already indicated that the expansion of our Universe is actually accelerating \cite{Perlmutter:1998np,Spergel:2003cb}. The investigation on how this accelerating phase could happen is still an active research direction currently. Typically, one possible direction to explain the accelerated expansion of the Universe is by introducing some sort of exotic matter fields called dark energy, with the simplest candidate being the cosmological constant $\Lambda$. Theoretically, dark energy is required to violate the strong energy condition, applying some sort of repulsive forces, such that it can drive the acceleration of the cosmic expansion. Another possible explanation for the accelerated expansion is by assuming that Einstein's general relativity (GR) has to be modified in some way. In this scenario, one can say that the left-hand side (geometry sector) of the Einstein equation is modified, while the right-hand side (matter sector) remains intact. The simplest modified gravitational theory following this direction is the $f(R)$ gravity, in which the Ricci scalar $R$ term in the Einstein-Hilbert action is replaced with an arbitrary function of it \cite{Sotiriou:2008rp,DeFelice:2010aj}. It should be emphasized that the scrutiny of modified theories of gravity can also be motivated in order to ameliorate the incompleteness of GR at high curvature regimes \cite{Capozziello:2011et}, such as the existence of spacetime singularities and its incompatibility with quantum mechanics. These issues require the formulation of a complete quantum theory of gravity. However, such a fundamental theory is still far beyond reach so far. Therefore, one may consider an extension of Einstein's GR such that the theory reduces to GR in proper limits and in the meantime resolves some problems of GR in the high curvature regimes. From a phenomenological perspective, these modified theories of gravity can also be treated as effective theories of the unknown quantum theory of gravity. For a more recent review on modified theories of gravity, we would like to refer the reader to Ref.~\cite{Nojiri:2017ncd}.      

In addition to changing either the left-hand side or the right-hand side of the Einstein equation, one can consider the possibility of modifying both the geometry and matter sectors simultaneously. This motivates the theories in which the matter fields are nonminimally coupled with the geometry sector. In fact, these theories could also explain the accelerated expansion of the Universe in an elegant manner. It should be noted that the $f(R)$ gravity is equivalent to a subclass of scalar-tensor theories in which there is a scalar field nonminimally coupled to gravity. In Refs.~\cite{Nojiri:2004bi,Allemandi:2005qs}, a gravitational theory in which the matter Lagrangian $\mathcal{L}_m$ is nonminimally coupled to curvature was proposed to explain the acceleration of the Universe. The Lagrangian of this model is given by $f_1(R)+f_2(R)\mathcal{L}_m$, and this model was discussed in more detail later in Ref.~\cite{Bertolami:2007gv}. Furthermore, in Ref.~\cite{Harko:2008qz}, the model of Ref.~\cite{Bertolami:2007gv} was extended by replacing the matter Lagrangian in the nonminimal coupling term with an arbitrary functional dependence. Furthermore, this theory was reformulated in the Palatini variational principle in Ref.~\cite{Harko:2010hw}. The most general functional dependence of the curvature $R$ and the matter Lagrangian $\mathcal{L}_m$, that is, the $f(R,\mathcal{L}_m)$ gravity, was proposed in Ref.~\cite{Harko:2010mv}.

Besides considering the nonminimal coupling between curvature and the matter Lagrangian, one can consider the nonminimal coupling between curvature and the energy-momentum tensor. Following this direction, in Ref.~\cite{Harko:2011kv}, the $f(R,T)$ theory was proposed, where $T$ is the trace of the energy-momentum tensor, and it has drawn a lot of attention recently. Also, the $f(R,T)$ gravity can be further generalized by introducing a new scalar invariant, which is defined by contracting the Ricci tensor and the energy-momentum tensor. The resultant theory is called the $f(R,T,R_{\mu\nu}T^{\mu\nu})$ gravity \cite{Haghani:2013oma,Odintsov:2013iba}. The $f(R,T)$ gravity and the $f(R,\mathcal{L}_m,R_{\mu\nu}T^{\mu\nu})$ have also been reformulated within the Palatini variational principle in Refs.~\cite{Wu:2018idg,Barrientos:2018cnx} and Ref.~\cite{Fox:2018gop}, respectively. It should be stressed that in all the theories mentioned above, because of the nonminimal couplings, the particles follow nongeodesic motion, and the equivalence principles are then violated.
 
In addition to the theories mentioned above, it is possible to consider a new scalar invariant, which contains nonlinear contributions of the energy-momentum tensor, to encode the nonminimal matter-geometry coupling. In Ref.~\cite{Arik:2013sti}, such an attempt was made for the first time. The Lagrangian contains an arbitrary functional of the Ricci scalar and the square of the energy-momentum tensor, that is, $f(R,T^{\mu\nu}T_{\mu\nu})$ gravity. In this so-called generalized energy-momentum squared gravity (gEMSG), the cosmological solutions have been studied in Refs.~\cite{Roshan:2016mbt,Akarsu:2017ohj,Board:2017ign,Keskin:2018bkg,Akarsu:2018aro,Akarsu:2019ygx,Faria:2019ejh,Bahamonde:2019urw}. It has been found that the nonlinear matter contributions in the field equations would affect the cosmological evolutions at high density regimes \cite{Roshan:2016mbt,Board:2017ign} and at low density regimes \cite{Akarsu:2017ohj}, depending on the specific form of the functional $f$ and the parameters under consideration. Some astrophysical applications, such as neutron stars, have also been studied in the context of energy-momentum squared gravity (EMSG) \cite{Akarsu:2018zxl,Nari:2018aqs}, in which $f(R,T^{\mu\nu}T_{\mu\nu})=R-\eta T^{\mu\nu}T_{\mu\nu}$ with $\eta$ being a constant.

In this paper, we are going to consider the black hole physics in the gEMSG. Presumably due to the fact that the nonlinear matter coupling does not alter the vacuum solutions, less attention has been paid to the black hole solutions in this model. In Ref.~\cite{Roshan:2016mbt}, the charged black hole solutions of the EMSG in which $f(R,T^{\mu\nu}T_{\mu\nu})=R-\eta T^{\mu\nu}T_{\mu\nu}$ are derived, and they deviate from the Reissner-Nordstr\"om (RN) black hole. In this paper, we will consider the gEMSG, without assuming any particular functional form of $f$ and study the axial perturbations of generic charged black holes in this type of theories. Generally speaking, when perturbing a black hole, the black hole undergoes a damping oscillation which is characterized by a superposition of many exponentially decaying sinusoidal modes, called quasinormal modes (QNMs) \cite{Chandrabook,Regge:1957td}. According to how they react with respect to the parity change, these perturbations can be divided into the axial perturbations and the polar perturbations. Because during the oscillations the black hole loses its energy by emitting gravitational waves, the whole system is a dissipative system, and that is the reason why the notion \textit{quasi} appears. Essentially, the QNMs have complex-valued frequencies. The real part describes the oscillation, and the imaginary part corresponds to the decay. In GR, the QNM frequencies satisfy the black hole no-hair theorem. Recently, inspired by the direct detection of gravitational waves from binary mergers \cite{Abbott:2016blz,Abbott:2017oio,TheLIGOScientific:2017qsa}, many attempts have been made to test black holes or gravitational theories by using the ringings of black holes \cite{Kobayashi:2012kh,Kobayashi:2014wsa,Minamitsuji:2014hha,Dong:2017toi,Tattersall:2018nve,Bhattacharyya:2017tyc,Bhattacharyya:2018qbe,Datta:2019npq,Chen:2018mkf,Chen:2018vuw,Blazquez-Salcedo:2016enn,Blazquez-Salcedo:2017txk,Blazquez-Salcedo:2018pxo,Toshmatov:2016bsb,Cardoso:2003qd,Cardoso:2003vt,Cardoso:2004cj,Chen:2019iuo,Volkel:2019muj,Bao:2019kgt}. This thus conceptualizes the field of \textit{black hole spectroscopy} \cite{Nollert:1999ji,Berti:2009kk,Konoplya:2011qq,Berti:2015itd,Cardoso:2019mqo,McManus:2019ulj,Glampedakis:2017dvb}.  

Theoretically, the QNMs and the perturbations are described by a set of master equations. To derive the master equations of the axial perturbations for the charged black holes in the gEMSG, we have to perturb the gravitational equations and the modified Maxwell equations in this theory. After deriving the master equations, in this work, we will focus on the modes in the eikonal approximation. It is well known that for most static, spherically symmetric, and asymptotically flat black holes in GR their eikonal QNMs can be directly determined by the properties of the photon ring of the black holes \cite{Cardoso:2008bp}. Such a correspondence has also been developed and investigated for rotating black holes \cite{Dolan:2010wr,Yang:2012he} and for situations with coupled master equations \cite{Glampedakis:2019dqh}. It would be interesting to see whether the correspondence between the eikonal QNMs and the black hole photon rings is still valid for other gravitational theories \cite{Churilova:2019jqx}. In fact, this correspondence has been found violated in the Einstein-Lovelock theory with a spacetime dimension $D>4$ \cite{Konoplya:2017lhs,Konoplya:2017wot}. Similar violation has also been found for the charged black holes in GR coupled with nonlinear electrodynamics \cite{Toshmatov:2018tyo,Toshmatov:2018ell} and for the charged black holes in the Eddington-inspired-Born-Infeld gravity \cite{Chen:2018vuw}. In the latter case, the nontrivial matter-geometry couplings inherent in the theory \cite{Banados:2010ix,Pani:2012qb,Delsate:2012ky} may play a crucial role in causing this violation. One can also draw similar conclusions when considering the perturbations of a scalar field nonminimally coupled to geometry \cite{Chen:2010qf}. In this paper, we will show explicitly that due to the nonminimal matter-geometry coupling in the gEMSG, the correspondence between the eikonal QNMs and the photon ring of the charged black holes ceases to be satisfied. It should be emphasized that we will not assume any specific functional form of $f(R,T^{\mu\nu}T_{\mu\nu})$ throughout this paper to ensure the validity of our conclusions to be as general as possible.

This paper is outlined as follows. In Sec.~\ref{sec.EOM}, we will briefly review the equations of motion of the gEMSG. We will also present the modified Klein-Gordon equation for a scalar field and the modified Maxwell equations for electromagnetic fields in this theory. In Sec.~\ref{sec.axial}, we will derive the coupled master equations of the axial perturbations for the charged black holes in the gEMSG. In Sec.~\ref{sec.eik}, we will show explicitly that the correspondence between the eikonal QNMs and the photon ring of the black hole is broken due to the nonminimal matter-geometry coupling. We finally draw our conclusions in Sec.~\ref{sec.conclu}.

\section{Equations of motion}\label{sec.EOM}
We start with the action of the gEMSG (we have assumed the speed of light $c=1$) \cite{Arik:2013sti},
\begin{equation}
\mathcal{S}=\frac{1}{2\kappa}\int\sqrt{-g}f(R,\bold{T}^2)d^4x+\mathcal{S}_m\,,\label{actiongeneral}
\end{equation}
where $\kappa=8\pi G$ and $\bold{T}^2\equiv T_{\alpha\beta}T^{\alpha\beta}$ is the square of the energy-momentum tensor. The function $f(R,\bold{T}^2)$ is an arbitrary function of the Ricci scalar $R$ and $\bold{T}^2$. Furthermore, $\mathcal{S}_m$ stands for the matter action. The equations of motion are derived by varying the action \eqref{actiongeneral}. The variation of $\bold{T}^2$ with respect to the metric $g_{\mu\nu}$ can be written as
\begin{align}
\delta_g\bold{T}^2&=\delta_g\left(g^{\alpha\rho}g^{\beta\sigma}T_{\alpha\beta}T_{\rho\sigma}\right)\nonumber\\
&=2\delta g^{\alpha\rho}{T_\alpha}^\sigma T_{\rho\sigma}+2T^{\alpha\beta}\delta_g T_{\alpha\beta}\nonumber\\
&=2\left({T_\mu}^\sigma T_{\nu\sigma}+\bold{\Psi}_{\mu\nu}\right)\delta g^{\mu\nu}\nonumber\\
&=\bold{\Theta}_{\mu\nu}\delta g^{\mu\nu}\,,\label{varyingTT}
\end{align}
where 
\begin{equation}
\bold{\Theta}_{\mu\nu}\equiv 2\left({T_\mu}^\sigma T_{\nu\sigma}+\bold{\Psi}_{\mu\nu}\right)\,,\quad\textrm{and}\quad \bold{\Psi}_{\mu\nu}\equiv T^{\alpha\beta}\frac{\delta T_{\alpha\beta}}{\delta g^{\mu\nu}}\,.\label{thetapsi}
\end{equation}
The subscript $g$ in $\delta_g$ is to emphasize that the variation is performed with respect to the metric $g_{\mu\nu}$. If the variation is done on objects which depend only on the metric, the subscript $g$ will be removed.

Using Eq.~\eqref{varyingTT}, the variation of the action \eqref{actiongeneral} with respect to $g_{\mu\nu}$ gives the field equation
\begin{equation}
f_R R_{\mu\nu}+\left(g_{\mu\nu}\Box-\nabla_\mu\nabla_\nu\right)f_R-\frac{f}{2}g_{\mu\nu}+f_{\bold{T}^2}\bold{\Theta}_{\mu\nu}=\kappa T_{\mu\nu}\,,\label{fieldeq}
\end{equation}
where $\Box=\nabla_\mu\nabla^\mu$ and we have defined $f_R\equiv\partial f/\partial R$ and $f_{\bold{T}^2}\equiv \partial f/\partial\bold{T}^2$. The energy-momentum tensor is defined via the matter action as follows:
\begin{equation}
T_{\mu\nu}\equiv-\frac{2}{\sqrt{-g}}\frac{\delta\mathcal{S}_m}{\delta g^{\mu\nu}}\,.\label{defiintionT}
\end{equation}

In the gEMSG, the standard conservation equation of the energy-momentum tensor is not satisfied; i.e., the covariant derivative of $T_{\mu\nu}$ does not vanish. This can be proven by taking a covariant derivative in Eq.~\eqref{fieldeq}. Combining it with the Bianchi identity $\nabla^\mu G_{\mu\nu}=0$, one can obtain \cite{Bahamonde:2019urw}
\begin{equation}
\kappa\nabla^\mu T_{\mu\nu}=-\frac{1}{2}f_{\bold{T}^2}\nabla_\nu\bold{T}^2+\nabla^\mu\left(f_{\bold{T}^2}\bold{\Theta}_{\mu\nu}\right)\,.\label{modifiedcons}
\end{equation}
Note that the identity $(\Box\nabla_\nu-\nabla_\nu\Box)f_R=R_{\mu\nu}\nabla^\mu f_R$ has been used. According to Eq.~\eqref{modifiedcons}, the standard conservation equation of the energy-momentum tensor is not fulfilled in this theory.  

\subsection{Modified Klein-Gordon equation}
As we have shown, the conservation equation of the energy-momentum tensor in the gEMSG acquires some modifications due to the nonminimal coupling of $\bold{T}^2$ in the gravitational action [see Eq.~\eqref{modifiedcons}]. Therefore, if we consider a scalar field or electromagnetic fields in curved spacetimes in this theory, the standard Klein-Gordon equation and the Maxwell equations will be modified. In this subsection, we will show how the Klein-Gordon equation is altered in the gEMSG. The modified Maxwell equations, on the other hand, will be presented later in the next subsection.

To investigate the modified Klein-Gordon equation, we consider the matter action of a scalar field $\psi$,
\begin{equation}
\mathcal{S}_m\equiv\mathcal{S}_\psi=\int d^4x\sqrt{-g}L_\psi\,,
\end{equation}
where
\begin{equation}
L_\psi=-\frac{1}{2}g^{\mu\nu}\partial_\mu\psi \partial_\nu\psi-V(\psi)\,,
\end{equation}
in which $V(\psi)$ is the scalar field potential. According to Eq.~\eqref{defiintionT}, the energy-momentum tensor of the scalar field can be written as
\begin{equation}
T^{(\psi)}_{\mu\nu}=\partial_\mu\psi\partial_\nu\psi+g_{\mu\nu}L_\psi\,.\label{Tphi}
\end{equation}

The modified Klein-Gordon equation is derived by varying the action \eqref{actiongeneral} with respect to the scalar field $\psi$:
\begin{align}
\delta_\psi\mathcal{S}=\int\sqrt{-g}d^4x \left(\frac{1}{2\kappa}f_{\bold{T}^2}\delta_\psi\bold{T}^2+\delta_\psi L_\psi\right)=0\,.\label{varyingscalarfield1}
\end{align}
The variation of $\bold{T}^2$ can be written as
\begin{align}
\delta_\psi\bold{T}^2&=2T^{(\psi)\mu\nu}\delta_\psi T^{(\psi)}_{\mu\nu}\nonumber\\
&=2T^{(\psi)\mu\nu}\left(2\partial_\mu\psi\delta\partial_\nu\psi+g_{\mu\nu}\delta_\psi L_\psi\right)\nonumber\\
&=4T^{(\psi)\mu\nu}\partial_\mu\psi\delta\partial_\nu\psi+2T^{(\psi)}\delta_\psi L_\psi\,.
\end{align}
Note that we have
\begin{equation}
\delta_\psi L_\psi=-g^{\mu\nu}\partial_\mu\psi\delta\partial_\nu\psi-\frac{dV}{d\psi}\delta\psi\,.
\end{equation}
After some algebra, one obtains the modified Klein-Gordon equation from Eq.~\eqref{varyingscalarfield1},
\begin{align}
&\partial_\nu\left[\sqrt{-g}\left(\frac{1}{\kappa}f_{\bold{T}^2}T^{(\psi)}+1\right)\partial^\nu\psi\right]\nonumber\\&-\frac{2}{\kappa}\partial_\nu\left(\sqrt{-g}f_{\bold{T}^2}T^{(\psi)\mu\nu}\partial_\mu\psi\right)\nonumber\\
=&\,\sqrt{-g}\left(\frac{1}{\kappa}f_{\bold{T}^2}T^{(\psi)}+1\right)\frac{dV}{d\psi}\,,
\end{align}
where $T^{(\psi)}$ is the trace of the energy-momentum tensor \eqref{Tphi}. The modified Klein-Gordon equation can also be rewritten as
\begin{widetext}
\begin{equation}
\Box\psi-\frac{dV}{d\psi}=\frac{1}{\kappa}\left[T^{(\psi)}f_{\bold{T}^2}\frac{dV}{d\psi}+2\nabla_\mu\left(f_{\bold{T}^2}T^{(\psi)\mu\nu}\nabla_\nu\psi\right)-\nabla_\mu\left(f_{\bold{T}^2}T^{(\psi)}\nabla^\mu\psi\right)\right]\,.\label{modifiedKG}
\end{equation}
\end{widetext}
It can be seen that if $f_{\bold{T}^2}=0$ the right-hand side of Eq.~\eqref{modifiedKG} vanishes, and the standard Klein-Gordon equation $\Box\psi-dV/d\psi=0$ is recovered. It should be mentioned that the modified Klein-Gordon equation \eqref{modifiedKG} can also be derived by inserting the energy-momentum tensor \eqref{Tphi} into the modified conservation equation \eqref{modifiedcons}.

\subsection{Modified Maxwell equations}
In this paper, we are going to study the perturbations of charged black holes in the gEMSG. In addition to perturbing the gravitational equation \eqref{fieldeq}, we have to perturb the modified Maxwell equations in this theory as well. Note that the modified Maxwell equations for the EMSG in which $f(R,\bold{T}^2)=R-\eta\bold{T}^2$ with a constant $\eta$ have been derived in order to study the charged black hole solutions in the theory \cite{Roshan:2016mbt}. In this subsection, we will derive the modified Maxwell equations in the gEMSG with the action \eqref{actiongeneral}. 

We consider the Lagrangian density of the electromagnetic fields, which reads
\begin{equation}
L_{em}=-\frac{1}{16\pi}\bold{F}^2\,,\qquad \bold{F}^2\equiv F_{\mu\nu}F^{\mu\nu}\,,\label{lagrangianemfield}
\end{equation}
where $F_{\mu\nu}=\partial_\mu A_\nu-\partial_\nu A_\mu$ is the field strength tensor. The energy-momentum tensor of the electromagnetic fields is given by
\begin{equation}
T_{\mu\nu}^{(em)}=\frac{1}{4\pi}\left(F_{\mu\sigma}{F_{\nu}}^\sigma-\frac{1}{4}g_{\mu\nu}\bold{F}^2\right)\,.\label{Tem}
\end{equation}

The equation of motion of the electromagnetic fields is derived by varying the action \eqref{actiongeneral} with respect to the vector potential $A_\mu$. We then have
\begin{equation}
\nabla_\mu\frac{\partial\left(8\pi f-\kappa\bold{F}^2\right)}{\partial\left(\nabla_\mu A_\nu\right)}=0\,.\label{maxwelleq1}
\end{equation}
To proceed, we use the identity
\begin{equation}
\frac{\partial F_{\alpha\beta}}{\partial\left(\nabla_\mu A_\nu\right)}=\delta^\mu_\alpha\delta^\nu_\beta-\delta^\mu_\beta\delta^\nu_\alpha
\end{equation}
to get
\begin{align}
&\frac{\partial\bold{F}^2}{\partial\left(\nabla_\mu A_\nu\right)}=4F^{\mu\nu}\,,\\
&\frac{\partial\bold{T}^2}{\partial\left(\nabla_\mu A_\nu\right)}=\frac{1}{2\pi^2}\left({F_\gamma}^\nu F^{\mu\rho}{F^\gamma}_\rho-\frac{1}{4}F^{\mu\nu}\bold{F}^2\right)\,.
\end{align}
As a result, Eq.~\eqref{maxwelleq1} gives the modified Maxwell equation
\begin{equation}
\nabla_\mu F^{\mu\nu}=\frac{1}{\kappa\pi}\nabla_\mu\left[f_{\bold{T}^2}\left({F_\gamma}^\nu F^{\mu\rho}{F^\gamma}_\rho-\frac{1}{4}F^{\mu\nu}\bold{F}^2\right)\right]\,.\label{modifiedMeq2}
\end{equation}
In addition, the field strength satisfies the Bianchi identity
\begin{equation}
\nabla_{[\mu}F_{\nu\lambda]}=0\,,\label{embianchi}
\end{equation}
which follows directly from the definition of the field strength. As one can see from Eq.~\eqref{modifiedMeq2}, the nonminimal coupling of $\bold{T}^2$ in the gravitational action modifies the standard Maxwell equation. If $f_{\bold{T}^2}$ is zero, the standard Maxwell equation $\nabla_\mu F^{\mu\nu}=0$ is recovered. It should also be mentioned that the modified Maxwell equation \eqref{modifiedMeq2} can be derived by inserting the energy-momentum tensor \eqref{Tem} and the Bianchi identity \eqref{embianchi} into the modified conservation equation \eqref{modifiedcons}. 

Before closing this section, we will write down the gravitational equation of the gEMSG in the presence of electromagnetic fields for the convenience of later use. To do this, we first consider the tensor $\bold{\Psi}_{\mu\nu}$ defined in Eq.~\eqref{thetapsi} and express it with the field strength. Recalling the definition of $\bold{F}^2$ given in Eq.~\eqref{lagrangianemfield}, the variation of $\bold{F}^2$ with respect to the metric reads
\begin{equation}
\frac{\delta\bold{F}^2}{\delta g^{\mu\nu}}=2{F_\mu}^\beta F_{\nu\beta}\,,
\end{equation}
with which one can obtain
\begin{align}
&\bold{\Psi}_{\mu\nu}=T^{(em)\alpha\beta}\frac{\delta T^{(em)}_{\alpha\beta}}{\delta g^{\mu\nu}}\nonumber\\
=&\,\frac{1}{4\pi}T^{(em)\alpha\beta}\left(F_{\mu\beta}F_{\nu\alpha}+\frac{1}{4}g_{\alpha\mu}g_{\beta\nu}\bold{F}^2-\frac{1}{2}g_{\alpha\beta}{F_\mu}^\sigma F_{\nu\sigma}\right)\,.\label{Psiem1}
\end{align}
Since the trace of the energy-momentum tensor vanishes, i.e., $T^{(em)}=0$, the last term of the second line in Eq.~\eqref{Psiem1} does not contribute. As a result, the tensor $\bold{\Psi}_{\mu\nu}$ can be written as
\begin{equation}
\bold{\Psi}_{\mu\nu}=\frac{1}{16\pi^2}\left[{F_\gamma}^\beta F^{\gamma\alpha} F_{\alpha\mu} F_{\beta\nu}-\frac{1}{16}g_{\mu\nu}\left(\bold{F}^2\right)^2\right]\,.\label{boldpsiex}
\end{equation}
On the other hand, the ${T_\mu}^\sigma T_{\nu\sigma}$ term appearing in Eq.~\eqref{thetapsi} can be expressed with the field strength as follows:
\begin{align}
&{{T^{(em)}}_\mu}^\sigma T^{(em)}_{\nu\sigma}\nonumber\\=&\,\frac{1}{16\pi^2}\left(2{F_\gamma}^\beta F^{\gamma\alpha} F_{\alpha\mu} F_{\beta\nu}-\frac{1}{2}{F^\rho}_\mu F_{\rho\nu}\bold{F}^2\right)-\bold{\Psi}_{\mu\nu}\,.\label{boldpsipre}
\end{align}
Note that after taking a trace of the above equation one can obtain the square of the energy-momentum tensor expressed with the field strength:
\begin{equation}
\bold{T}^2={F^\alpha}_\theta F_{\alpha\rho} F^{\gamma\theta} {F_\gamma}^\rho-\frac{1}{4}\left(\bold{F}^2\right)^2\,.
\end{equation}
Finally, combining Eqs.~\eqref{thetapsi}, \eqref{Tem}, \eqref{boldpsiex}, and \eqref{boldpsipre}, the gravitational equation \eqref{fieldeq} can be written as
\begin{align}
f_RR_{\mu\nu}&+\left(g_{\mu\nu}\Box-\nabla_\mu\nabla_\nu\right)f_R-\frac{f}{2}g_{\mu\nu}\nonumber\\=&\,\frac{\kappa}{4\pi}\left(F_{\mu\sigma}{F_{\nu}}^\sigma-\frac{1}{4}g_{\mu\nu}\bold{F}^2\right)\nonumber\\&-\frac{f_{\bold{T}^2}}{8\pi^2}\left(2{F_\gamma}^\beta F^{\gamma\alpha} F_{\alpha\mu} F_{\beta\nu}-\frac{1}{2}{F^\rho}_\mu F_{\rho\nu}\bold{F}^2\right)\,.\label{fTem}
\end{align}

\section{Axial perturbations}\label{sec.axial}
To study the QNMs of the black holes, we consider the perturbations of a static and spherically symmetric spacetime. In the gEMSG, the black hole solutions can be derived only when an explicit expression of $f(R,\bold{T}^2)$ is given and the matter fields are specified. In this paper, we will consider electromagnetic fields and focus on the eikonal QNMs of charged black holes in the gEMSG, without assuming any specific expression of $f(R,\bold{T}^2)$. As we have mentioned in the Introduction, instead of calculating the QNM frequencies explicitly from some certain black hole solutions, we will derive the master equations of the QNMs and express them by using unspecified metric functions. Since the gravitational fields and electromagnetic fields for charged black holes are coupled together, the master equations describing QNMs have to be derived by perturbing the modified Maxwell equations [Eqs.~\eqref{modifiedMeq2} and \eqref{embianchi}] and the gravitational equation \eqref{fTem}. The two master equations turn out to couple with each other. Then, we will prove that the correspondence between the eikonal QNMs and the photon ring of the black hole would be violated in this theory. All these can indeed be done, even though the explicit expressions of the metric functions remain unspecified, and our conclusions turn out to be very general because they are valid for charged black holes in the gEMSG with an arbitrary $f(R,\bold{T}^2)$.

 Without loss of generality, the perturbed spacetime can be described by a nonstationary and axisymmetric metric in which the symmetrical axis is turned in such a way that no $\phi$ dependence appears in the metric functions. In general, the metric can be written as \cite{Chandrabook}
\begin{align}
ds^2=&-e^{2\nu}\left(dx^0\right)^2+e^{2\mu_1}\left(dx^1-\sigma dx^0-q_2dx^2-q_3dx^3\right)^2\nonumber\\&+e^{2\mu_2}\left(dx^2\right)^2+e^{2\mu_3}\left(dx^3\right)^2\,,\label{metricg}
\end{align}
where $\nu$, $\mu_1$, $\mu_2$, $\mu_3$, $\sigma$, $q_2$, and $q_3$ are functions of time $t$ ($t=x^0$), radial coordinate $r$ ($r=x^2$), and polar angle $\theta$ ($\theta=x^3$). Because the system is axisymmetric, the metric functions are independent of the azimuthal angle $\phi$ ($\phi=x^1$). Note that in the background spacetime which is static and spherically symmetric, we have $\sigma=q_2=q_3=0$.

\subsection{Tetrad formalism}

To study the perturbations of the spacetime metric \eqref{metricg}, we will use the tetrad formalism in which one defines a basis associated with the metric \eqref{metricg} \cite{Chandrabook},
\begin{align}
e^{\mu}_{(0)}&=\left(e^{-\nu},\quad\sigma e^{-\nu},\quad0,\quad0\right)\,,\nonumber\\
e^{\mu}_{(1)}&=\left(0,\quad e^{-\mu_1},\quad 0,\quad0\right)\,,\nonumber\\
e^{\mu}_{(2)}&=\left(0,\quad q_2e^{-\mu_2},\quad e^{-\mu_2},\quad0\right)\,,\nonumber\\
e^{\mu}_{(3)}&=\left(0,\quad q_3e^{-\mu_3},\quad 0,\quad e^{-\mu_3}\right)\,,\label{tetradbasis111}
\end{align}
and
\begin{align}
e_{\mu}^{(0)}&=\left(e^{\nu},\quad0,\quad0,\quad0\right)\,,\nonumber\\
e_{\mu}^{(1)}&=\left(-\sigma e^{\mu_1},\quad e^{\mu_1},\quad -q_2e^{\mu_1},\quad -q_3e^{\mu_1}\right)\,,\nonumber\\
e_{\mu}^{(2)}&=\left(0,\quad 0,\quad e^{\mu_2},\quad0\right)\,,\nonumber\\
e_{\mu}^{(3)}&=\left(0,\quad 0,\quad 0,\quad e^{\mu_3}\right)\,,\label{tetradbasis222}
\end{align}
where the tetrad indices are enclosed in parentheses to distinguish them from the tensor indices. The tetrad basis should satisfy
\begin{align}
e_{\mu}^{(a)}e^{\mu}_{(b)}&=\delta^{(a)}_{(b)}\,,\quad e_{\mu}^{(a)}e^{\nu}_{(a)}=\delta^{\nu}_{\mu}\,,\nonumber\\
e_{\mu}^{(a)}&=g_{\mu\nu}\eta^{(a)(b)}e^{\nu}_{(b)}\,,\nonumber\\
g_{\mu\nu}&=\eta_{(a)(b)}e_{\mu}^{(a)}e_{\nu}^{(b)}\equiv e_{(a)\mu}e_{\nu}^{(a)}\,.
\end{align}
Conceptually, in the tetrad formalism, we project the relevant quantities defined on the coordinate basis of $g_{\mu\nu}$ onto a chosen basis of $\eta_{(a)(b)}$ by constructing the tetrad basis correspondingly. In practice, $\eta_{(a)(b)}$ is usually assumed to be the Minkowskian matrix
\begin{equation}
\eta_{(a)(b)}=\eta^{(a)(b)}=\textrm{diag}\left(-1,1,1,1\right)\,.
\end{equation}
In this regard, any vector or tensor field can be projected onto the tetrad frame in which the field can be expressed through its tetrad components:
\begin{align}
A_{\mu}&=e_{\mu}^{(a)}A_{(a)}\,,\quad A_{(a)}=e_{(a)}^{\mu}A_{\mu}\,,\nonumber\\
B_{\mu\nu}&=e_{\mu}^{(a)}e_{\nu}^{(b)}B_{(a)(b)}\,,\quad B_{(a)(b)}=e_{(a)}^{\mu}e_{(b)}^{\nu}B_{\mu\nu}\,.
\end{align}
One should notice that in the tetrad formalism the covariant (partial) derivative in the original coordinate frame is replaced with the intrinsic (directional) derivative in the tetrad frame. For instance, the derivatives of an arbitrary rank-2 object $H_{\mu\nu}$ in the two frames are related as \cite{Chandrabook}
\begin{align}
&\,H_{(a)(b)|(c)}\equiv e^{\lambda}_{(c)}H_{\mu\nu;\lambda}e_{(a)}^{\mu}e_{(b)}^{\nu}\nonumber\\
=&\,H_{(a)(b),(c)}\nonumber\\&-\eta^{(m)(n)}\left(\gamma_{(n)(a)(c)}H_{(m)(b)}+\gamma_{(n)(b)(c)}H_{(a)(m)}\right)\,,\label{2.7}
\end{align}
where a vertical rule and a comma denote the intrinsic and directional derivatives with respect to the tetrad indices, respectively. A semicolon denotes a covariant derivative with respect to the tensor indices. Furthermore, the Ricci rotation coefficients are defined by
\begin{equation}
\gamma_{(c)(a)(b)}\equiv e_{(b)}^{\mu}e_{(a)\nu;\mu}e_{(c)}^{\nu}\,,
\end{equation}
and their components corresponding to the metric \eqref{metricg} are given in Ref.~\cite{Chandrabook}.

\subsection{Perturbed Maxwell equations in gEMSG}
Now, we will perturb the modified Maxwell equations in the gEMSG. For the background level, we consider the field strength with a purely radial electric field and no magnetic field. Therefore, only the $(t,r)$ and $(r,t)$ components, i.e., the $(0,2)$ and $(2,0)$ components of the field strength $F_{\mu\nu}$, appear at the background level.

We then linearize the equations by decomposing, for instance, $F_{(0)(2)}$ as $F_{(0)(2)}\rightarrow F_{(0)(2)}+\delta F_{(0)(2)}$ in which $F_{(0)(2)}$ is the field strength at the background level.{\footnote{We only use a delta to express the linear-order perturbations of quantities whose values at the background level do not vanish. The quantities that vanish at the background level, such as the metric functions $\sigma$, $q_2$, $q_3$, and the Maxwell tensor components $F_{(i)(j)}$ ($ij\ne02$ or $20$), shall be regarded as linear-order perturbation quantities directly.}} The linearized Maxwell equation \eqref{modifiedMeq2} gives
\begin{align}
\left(re^\nu X_{(1)(2)}\right)_{,r}&+e^{\nu+\mu_2}X_{(1)(3),\theta}+re^{\mu_2}X_{(0)(1),t}\nonumber\\&=r^2\sin{\theta}X_{(0)(2)}\left(\sigma_{,r}-q_{2,t}\right)\,,\label{17}
\end{align}
where $\bold{F}^2=-2F_{(0)(2)}^2$. In the above expression, we have defined a tensor field $X_{\mu\nu}$ whose tetrad components read
\begin{align}
X_{(a)(b)}&=F_{(a)(b)}\nonumber\\&-\frac{f_{\bold{T}^2}}{\kappa\pi}\left(F_{(c)(b)}{F_{(a)}}^{(d)}{F^{(c)}}_{(d)}-\frac{1}{4}F_{(a)(b)}\bold{F}^2\right)\,.\label{XFREL}
\end{align}
Using Eq.~\eqref{XFREL}, the axial components of the perturbed $X_{(a)(b)}$ can be expressed with those of the perturbed $F_{(a)(b)}$ as
\begin{align}
X_{(1)(2)}&=F_{(1)(2)}\sigma_+\,,\quad X_{(1)(3)}=F_{(1)(3)}\sigma_-\,,\nonumber\\ &X_{(0)(1)}=F_{(0)(1)}\sigma_+\,,
\end{align}
where
\begin{equation}
\sigma_\pm\equiv 1\pm\frac{f_{\bold{T}^2}}{2\kappa\pi}F_{(0)(2)}^2
\end{equation}
and $\sigma_\pm$ depends only on the radial coordinate $r$. The $X_{(0)(2)}$ in Eq.~\eqref{17} is a zero order quantity, and it can be solved to get
\begin{equation}
X_{(0)(2)}=F_{(0)(2)}\sigma_+=\frac{Q_*}{r^2}\,,\label{background02}
\end{equation}
where $Q_*$ is an integration constant, standing for the charge of the black hole.

On the other hand, the linearized Bianchi identity of the electromagnetic fields $F_{[(a)(b)|(c)]}$ gives
\begin{align}
&\left(re^\nu\sin\theta F_{(0)(1)}\right)_{,r}+re^{\mu_2}\sin\theta F_{(1)(2),t}=0\,,\label{45}\\
&re^\nu\left(F_{(0)(1)}\sin\theta\right)_{,\theta}+r^2\sin\theta F_{(1)(3),t}=0\label{46}\,.
\end{align}
For the sake of abbreviation, we define the following field perturbation:
\begin{equation}
B\equiv F_{(0)(1)}\sin\theta\,.
\end{equation}
After taking a derivative of Eq.~\eqref{17} with respect to $t$ and using Eqs.~\eqref{45} and \eqref{46}, we get
\begin{align}
\left[e^{\nu-\mu_2}\sigma_+\left(re^\nu B\right)_{,r}\right]_{,r}&+\frac{e^{2\nu+\mu_2}\sigma_-}{r}\left(\frac{B_{,\theta}}{\sin\theta}\right)_{,\theta}\sin\theta\nonumber\\
&-re^{\mu_2}\sigma_+ B_{,tt}\nonumber\\=-r^2\sigma_+&F_{(0)(2)}\left(\sigma_{,rt}-q_{2,tt}\right)\sin^2\theta\,.\label{48}
\end{align}
The left-hand side of this equation contains the axial perturbations of the matter field $B$. The right-hand side, on the other hand, contains the perturbations of the metric. Therefore, Eq.~\eqref{48} will become one of the coupled master equations.

\subsection{Perturbed gravitational equation}
Now, we consider the axial perturbations of the gravitational equation \eqref{fTem}. In the tetrad frame, this equation can be written as
\begin{widetext}
\begin{align}
f_RR_{(a)(b)}&+\eta_{(a)(b)}\left(\Box f_R-\frac{f}{2}\right)-e_{(a)}^\mu\left(f_{R,(b)}\right)_{,\mu}+\gamma_{(c)(b)(a)}f_{R,(d)}\eta^{(c)(d)}\nonumber\\=&\,\frac{\kappa}{4\pi}\left(F_{(a)(c)}{F_{(b)}}^{(c)}-\frac{1}{4}\eta_{(a)(b)}\bold{F}^2\right)-\frac{f_{\bold{T}^2}}{8\pi^2}\left(2{F_{(e)}}^{(d)} F^{(e)(c)} F_{(c)(a)} F_{(d)(b)}-\frac{1}{2}{F^{(c)}}_{(a)} F_{(c)(b)}\bold{F}^2\right)\,.
\end{align}
\end{widetext}
The $(1,3)$ and $(1,2)$ components of the perturbed gravitational equation give
\begin{align}
f_RR_{(1)(3)}+\gamma_{(2)(3)(1)}e^{-\mu_2}f_{R,r}&=0\,,\\
f_RR_{(1)(2)}+\frac{\kappa}{4\pi}\sigma_+F_{(0)(2)}F_{(0)(1)}&=0\,,
\end{align}
which can be written explicitly as
\begin{align}
\left[f_Rr^2e^{\nu-\mu_2}\left(q_{2,\theta}-q_{3,r}\right)\right]_{,r}&\nonumber\\-f_Rr^2&e^{-\nu+\mu_2}\left(\sigma_{,\theta}-q_{3,t}\right)_{,t}=0\,,\label{51}\\
\left[f_Rr^2e^{\nu-\mu_2}\left(q_{3,r}-q_{2,\theta}\right)\sin^3\theta\right]_{,\theta}&\nonumber\\-f_Rr^4e^{-\nu-\mu_2}\left(\sigma_{,r}-q_{2,t}\right)_{,t}&\sin^3\theta\nonumber\\=\frac{\kappa}{2\pi}\sigma_+&F_{(0)(2)}r^3e^\nu B\sin\theta\,,\label{52}
\end{align}
respectively. Note that in a static and spherically symmetric background $f_R$ depends only on the radial coordinate $r$. Then, we define
\begin{equation}
Q\equiv f_Rr^2 e^{\nu-\mu_2}\left(q_{2,\theta}-q_{3,r}\right)\sin^3\theta\,,
\end{equation}
with which Eqs.~\eqref{51} and \eqref{52} can be rewritten as
\begin{align}
e^{\nu-\mu_2}\frac{Q_{,r}}{f_Rr^2\sin^3\theta}&=\left(\sigma_{,\theta}-q_{3,t}\right)_{,t}\,,\label{54}\\
e^{\nu+\mu_2}\frac{Q_{,\theta}}{f_Rr^4\sin^3\theta}&\nonumber\\=-\left(\sigma_{,r}-q_{2,t}\right)_{,t}&-e^{2\nu+\mu_2}\frac{\kappa\sigma_+F_{(0)(2)}}{2\pi f_R r\sin^2\theta}B\,,\label{55}
\end{align}
respectively. By differentiating Eqs.~\eqref{54} and \eqref{55} and eliminating $\sigma$, we get
\begin{align}
&\frac{1}{\sin^3\theta}\left(\frac{e^{\nu-\mu_2}}{f_Rr^2}Q_{,r}\right)_{,r}+\frac{e^{\nu+\mu_2}}{f_Rr^4}\left(\frac{Q_{,\theta}}{\sin^3\theta}\right)_{,\theta}\nonumber\\=&\,\frac{e^{-\nu+\mu_2}}{f_Rr^2\sin^3\theta}Q_{,tt}-\frac{\kappa e^{2\nu+\mu_2}\sigma_+F_{(0)(2)}}{2\pi f_R r}\left(\frac{B}{\sin^2\theta}\right)_{,\theta}\,.\label{56}
\end{align}
As one can see, the left-hand side of this equation contains the metric perturbations, while the right-hand side of it contains the matter field perturbation $B$. Therefore, Eq.~\eqref{56} will become the second coupled master equation.

Furthermore, the right-hand side of Eq.~\eqref{48} can be expressed in terms of $Q$ as
\begin{align}
\left[e^{\nu-\mu_2}\sigma_+\left(re^\nu B\right)_{,r}\right]_{,r}&+\frac{e^{2\nu+\mu_2}\sigma_-}{r}\left(\frac{B_{,\theta}}{\sin\theta}\right)_{,\theta}\sin\theta\nonumber\\-re^{\mu_2}\sigma_+ B_{,tt}&-\frac{\kappa r}{2\pi f_R}e^{2\nu+\mu_2}\left(\sigma_+F_{(0)(2)}\right)^2B\nonumber\\&=\frac{e^{\nu+\mu_2}\sigma_+F_{(0)(2)}}{f_Rr^2\sin\theta}Q_{,\theta}\,,\label{57}
\end{align}
by using Eq.~\eqref{55}. Finally, Eqs.~\eqref{56} and \eqref{57} form two coupled master equations which describe the axial perturbations $Q$ and $B$. 

\subsection{Effective potentials}
For later convenience, we will recast the master equations \eqref{56} and \eqref{57} into a Schr\"odinger-like form with an effective potential in each equation. Essentially, the effective potential plays an important role in determining the properties of QNMs, including their behaviors in the eikonal limit.

We introduce the ansatz \cite{Chandrabook}
\begin{equation}
Q(r,\theta)=Q(r)Y(\theta)\,,\qquad B(r,\theta)=B(r)Y_{,\theta}/\sin\theta\,,
\end{equation}
where $Y(\theta)$ is the Gegenbauer function satisfying \cite{Abramow}
\begin{align}
\left(\frac{Y_{,\theta}}{\sin^3\theta}\right)_{,\theta}&=-\mu^2\frac{Y}{\sin^3\theta}\,,\\
\sin\theta\left[\frac{1}{\sin\theta}\left(\frac{Y_{,\theta}}{\sin\theta}\right)_{,\theta}\right]_{,\theta}&=-\left(\mu^2+2\right)\frac{Y_{,\theta}}{\sin\theta}\,,
\end{align}
where $\mu$ is related to the multipole number $l$ via $\mu^2=(l-1)(l+2)$. With this ansatz, Eqs.~\eqref{57} and \eqref{56} can be rewritten as
\begin{align}
&\left[e^{\nu-\mu_2}\sigma_+\left(re^\nu B\right)_{,r}\right]_{,r}\nonumber\\&+\left[\omega^2r\sigma_+e^{\mu_2}-\left(\mu^2+2\right)\frac{e^{2\nu+\mu_2}\sigma_-}{r}-\frac{4Q_*^2}{f_Rr^3}e^{2\nu+\mu_2}\right]B\nonumber\\
&=\frac{e^{\nu+\mu_2}Q_*}{f_Rr^4}Q\,,\label{61}\\
&\left(\frac{e^{\nu-\mu_2}}{f_Rr^2}Q_{,r}\right)_{,r}+\left(\frac{e^{-\nu+\mu_2}\omega^2}{f_Rr^2}-\frac{e^{\nu+\mu_2}\mu^2}{f_Rr^4}\right)Q\nonumber\\&=\frac{4e^{2\nu+\mu_2}Q_*\mu^2}{f_Rr^3}B\,,\label{62}
\end{align}
respectively, where we have used Eq.~\eqref{background02} and the Fourier decomposition $\partial_t\rightarrow-i\omega$. Note also that we have adopted the geometric unit system: $\kappa=8\pi$.

To proceed, we introduce the definitions
\begin{equation}
H_1^{(-)}\equiv -2\mu\sigma_+^{1/2}re^\nu B\,,\qquad H_2^{(-)}\equiv \frac{Q}{Z}\,,
\end{equation}
where $Z\equiv f_R^{1/2}r$. We also introduce the tortoise radius $r_*$, which is defined by
\begin{equation}
\frac{dr}{dr_*}=e^{\nu-\mu_2}\,.
\end{equation}
The master equations \eqref{61} and \eqref{62} can be rewritten as
\begin{align}
&\frac{d^2H_1^{(-)}}{dr_*^2}+\omega^2H_1^{(-)}\nonumber\\
=&\,\left[\frac{1}{2\sigma_+^{1/2}}\left(\frac{\sigma_{+,r_*}}{\sigma_+^{1/2}}\right)_{,r_*}+\left(\mu^2+2\right)\frac{e^{2\nu}}{r^2}\bold{\Gamma}+\frac{4Q_*^2e^{2\nu}}{f_Rr^4\sigma_+}\right]H_1^{(-)}\nonumber\\&-\frac{2\mu e^{2\nu}Q_*}{r^3\sqrt{f_R\sigma_+}}H_2^{(-)}\,,\label{fr222}\\
&\frac{d^2H_2^{(-)}}{dr_*^2}+\omega^2H_2^{(-)}\nonumber\\=&\,\left[-Z\left(\frac{Z_{,r_*}}{Z^2}\right)_{,r_*}+\frac{e^{2\nu}\mu^2}{r^2}\right]H_2^{(-)}-\frac{2\mu e^{2\nu}Q_*}{r^3\sqrt{f_R\sigma_+}}H_1^{(-)}\,,\label{fr111}
\end{align}
where $\bold{\Gamma}\equiv\sigma_-/\sigma_+$.

It can be seen that the coupled master equations have been recast into a Schr\"odinger-like form, and they can be written in a matrix expression as
\begin{equation}
\left(\frac{d^2}{dr_*^2}+\omega^2\right)
\begin{bmatrix}
    H_1^{(-)}  \\
    H_2^{(-)}
    \end{bmatrix}=
    \begin{bmatrix}
    V_{11} &V_{12} \\
    V_{21} &V_{22}
    \end{bmatrix}
    \begin{bmatrix}
    H_1^{(-)}  \\
    H_2^{(-)}
    \end{bmatrix}\,,\label{coupledeq1}
\end{equation}
where $V_{ij}$ is given by Eqs.~\eqref{fr222} and \eqref{fr111}.

According to the coupled master equations \eqref{fr222} and \eqref{fr111}, one can see that:
\begin{enumerate}[(i)]
\item If $f(R,\bold{T}^2)=R$ such that $f_R=1$ and $f_{\bold{T}^2}=0$, the theory reduces to GR: 
\begin{align}
V_{12}&=V_{21}=-\frac{2Q_*\mu}{r^3}e^{2\nu}\,,\\
V_{11}&=\frac{e^{2\nu}}{r^3}\left[(\mu^2+2)r+\frac{4Q_*^2}{r}\right]\,,\\
V_{22}&=\frac{e^{2\nu}}{r^3}\left[(\mu^2+2)r-3+\frac{4Q_*^2}{r}\right]\,,
\end{align}
where 
\begin{equation}
e^{2\nu}=1-\frac{1}{r}+\frac{Q_*^2}{r^2}\,.
\end{equation}
Therefore, the master equations reduce to those of the RN black hole as expected \cite{Chandrabook}.
\item If $f(R,\bold{T}^2)=R$ and $Q_*=0$, we have $V_{12}=V_{21}=0$ and
\begin{align}
V_{11}&=\frac{e^{2\nu}}{r^2}l(l+1)\,,\label{v11sw}\\
V_{22}&=\frac{e^{2\nu}}{r^2}\left[l(l+1)-\frac{3}{r}\right]\,.\label{v22sw}
\end{align}
Therefore, the effective potential for pure electromagnetic perturbations and for pure axial gravitational perturbations of the Schwarzschild black hole (the Regge-Wheeler equation \cite{Regge:1957td}) are recovered, respectively.
\end{enumerate}

\section{Eikonal QNMs and photon ring}\label{sec.eik}
As has been mentioned, the QNMs of the axial perturbations for the charged black holes in the gEMSG are described by the coupled master equations \eqref{fr222} and \eqref{fr111}. Clearly, the QNM spectrum depends on the value of the multipole number $l$. In this section, we will focus on the QNMs within the eikonal approximation and discuss the validity of their relation with the photon ring of the black hole.

For a simple illustration, we consider a single wave equation describing perturbations (say, $\Psi$ field) around a static and spherically symmetric black hole,
\begin{equation}
\left(\frac{d^2}{dr_*^2}+\omega^2\right)\Psi=V(r)\Psi\,,
\end{equation}
where $r_*$ is the tortoise radius defined in a standard way. If we assume that the black hole is asymptotically flat, the potential $V(r)$ is required to vanish when $r_*\rightarrow\pm\infty$ and has only one single peak at a finite $r_*$. In the study of QNMs, the above wave equation is solved by imposing the boundary conditions that only outgoing waves exist at spatial infinity ($r_*\rightarrow\infty$), while only incoming waves exist on the event horizon ($r_*\rightarrow-\infty$). In the eikonal optics prescription, it has been shown explicitly in Ref.~\cite{Glampedakis:2019dqh} that the real part and the imaginary part of the fundamental QNM frequency ($\omega=\omega_R+i\omega_I$) can be obtained by considering the leading order and the subleading order of the eikonal approximation, respectively \cite{Glampedakis:2019dqh},
\begin{equation}
\omega_R=\sqrt{V_p}\,,\qquad\omega_I=-\frac{1}{2}\sqrt{\frac{-V''_p}{2V_p}}\,,\label{eikonal1}
\end{equation} 
where the prime denotes the derivative with respect to $r_*$ and the index $p$ denotes the quantities evaluated at the peak of the potential. It should be emphasized that this result, Eq.~\eqref{eikonal1}, can also be derived by taking $l\rightarrow\infty$ in the standard WKB formula \cite{Schutz:1985zz,Iyer:1986np,Konoplya:2003ii,Matyjasek:2017psv,Konoplya:2019hlu}. In fact, according to the WKB formula, the imaginary part of the eikonal QNM frequency depends on the overtone number $n$ and the QNM frequencies for different overtones can be written as follows:
\begin{equation}
\omega_R=\sqrt{V_p}\,,\qquad\omega_I=-\left(n+\frac{1}{2}\right)\sqrt{\frac{-V''_p}{2V_p}}\,.\label{eikonal2}
\end{equation}
See also the following pioneer works \cite{Mashhoon:1982im,Ferrari:1984zz} in which a similar analytic expression for the QNM frequencies was deduced.

The most important consequence regarding the eikonal QNMs is their correspondence with the properties of the photon ring of the black hole. In Ref.~\cite{Cardoso:2008bp}, it is explicitly pointed out that for most cases of a static, spherically symmetric, and asymptotically flat black hole the eikonal QNM frequencies are related to the photon ring of the black hole. More precisely, the real part of the eikonal QNM frequencies corresponds to the angular frequency $\Omega_c$ of photons on the photon ring, and the imaginary part corresponds to the Lyapunov exponent $\lambda_c$ quantifying the instability of this circular orbit:
\begin{equation}
\omega=\Omega_cl-i\left(n+\frac{1}{2}\right)\left|\lambda_c\right|\,.\label{eikonalcorre}
\end{equation}     
The reason behind this correspondence is that for this class of black holes and master equations the potential $V$ in the eikonal limit can be approximated as
\begin{equation}
V(r)\approx \frac{e^{2\nu}}{r^2}l^2\,.\label{eikonalpotential}
\end{equation}
It turns out that the peak of this potential is located exactly on the photon ring. After inserting this potential into Eq.~\eqref{eikonal2}, one obtains Eq.~\eqref{eikonalcorre}.

Let us consider the master equations \eqref{fr222} and \eqref{fr111} in the gEMSG. The two master equations couple together in this case. However, as has been pointed in Ref.~\cite{Glampedakis:2019dqh}, as long as the coupled master equations satisfy the ``weak $l$ coupling" in which the term $V_{12}V_{21}$ does not appear in the leading-order eikonal approximation, the system can be treated as a decoupled system, and the result given in Eq.~\eqref{eikonal2} is still valid. Apparently, the term $V_{12}V_{21}$ derived from the master equations \eqref{fr222} and \eqref{fr111} only contains the multipole number $l$ up to quadratic order. Therefore, the weak $l$ coupling condition is satisfied. Furthermore, one can see that in the eikonal approximation, the potential $V_{22}$ given in Eq.~\eqref{fr111} reduces to that in Eq.~\eqref{eikonalpotential}; hence, the correspondence between the eikonal QNMs of the field $H_2^{(-)}$ and the photon ring of the charged black hole still holds. However, the potential $V_{11}$ given in Eq.~\eqref{fr222} does not reduce to that in Eq.~\eqref{eikonalpotential} because of the $\bold\Gamma$ factor. In this regard, as long as $\bold\Gamma\ne1$, the two perturbations $H_1^{(-)}$ and $H_2^{(-)}$ propagate independently, and they do not share the same eikonal frequencies. The eikonal QNMs for the field $H_1^{(-)}$ can be described by Eq.~\eqref{eikonalcorre} only when $\sigma_\pm=1$, that is, either in the absence of charge $Q_*$, or when the nonminimal coupling between matter and geometry is turned off ($f_{\bold{T}^2}=0$).

Before closing this section, we would like to mention that the correspondence between the eikonal QNMs and the photon ring of the black hole, i.e., Eq.~\eqref{eikonalcorre}, has been found to be not satisfied in some particular models. For example, for the higher-dimensional black holes in the Lovelock gravity, this correspondence is not fulfilled in general \cite{Konoplya:2017lhs,Konoplya:2017wot}. Similar violation can be found even in GR when considering nonlinear electrodynamics \cite{Toshmatov:2018tyo,Toshmatov:2018ell}. Recently, a similar violation has been pointed out for the charged black holes in the Eddington-inspired-Born-Infeld gravity \cite{Chen:2018vuw}, and it may be due to the nontrivial matter-geometry coupling inherent in the theory. The result in this paper tends to support this argument in the sense that the nonminimal matter coupling in the gEMSG also breaks the correspondence between the eikonal QNMs and the photon ring of the black hole.

\section{Conclusions}\label{sec.conclu}
In this paper, we consider the charged black holes in the gEMSG and study their axial perturbations. The action of the theory is constructed by an arbitrary function of the Ricci scalar and the square of the energy-momentum tensor, i.e., $f(R,\bold{T}^2)$, and therefore the theory contains nonminimal couplings between matter and geometry. Because of the nonminimal matter-geometry coupling in the theory, the energy-momentum tensor is not as conserved as that in GR. Without assuming any explicit expression for the function $f$, we first derive the modified Klein-Gordon equation for a scalar field and the modified Maxwell equations for electromagnetic fields in this model. Then, we perturb the gravitational equation and the modified Maxwell equations to obtain the coupled master equations of the axial perturbations for the charged black holes, that is, Eqs.~\eqref{fr222} and \eqref{fr111}.

After obtaining the master equations of the axial perturbations, we focus on the QNMs within the eikonal approximation. It is well known that for most black hole solutions in GR the QNM frequencies in the eikonal limit can be determined by the angular frequency and the Lyapunov exponent of the photon ring of the black hole. This correspondence is due to the fact that the peak of the effective potential in the master equations is exactly on the photon ring. In the gEMSG, however, this correspondence is not satisfied because of the nonminimal matter-geometry coupling [the $\bold{\Gamma}$ factor in Eq.~\eqref{fr222}]. In fact, the two coupled fields of the axial perturbations propagate independently, and they do not share the same QNM frequencies.

In Ref.~\cite{Chen:2018vuw}, it has been shown that the correspondence between the eikonal QNMs and the photon ring of the black holes is similarly broken for the charged black holes in the Eddington-inspired-Born-Infeld gravity. In fact, one can recast such a model into the Einstein frame, and a nontrivial matter-geometry coupling naturally appears \cite{Delsate:2012ky}. In this paper, we have provided an example in which the correspondence between the eikonal QNMs and the photon ring ceases to be valid in the presence of nonminimal matter-geometry couplings. This work can be extended by considering a gravitational theory with a more general nonminimal matter-geometry coupling, whose action may be written as
\begin{equation}
\mathcal{S}=\frac{1}{2\kappa}\int\sqrt{-g}f(R,\bold{T}^2,T,R_{\mu\nu}T^{\mu\nu},\mathcal{L}_m)d^4x+\mathcal{S}_m\,,\label{actiongeneralmost}
\end{equation}
where $T\equiv g^{\mu\nu}T_{\mu\nu}$ is the trace of the energy-momentum tensor and $\mathcal{L}_m$ stands for the matter Lagrangian.{\footnote{Since $T=0$ for Maxwell electromagnetic fields, it can be expected that there is no nonminimal matter-geometry coupling contributed from the presence of $T$ in the action when considering Maxwell electromagnetic fields in the matter sector. The reason of adding $T$ into Eq.~\eqref{actiongeneralmost} is just for the sake of generality.}} It will be interesting to see how the correspondence between the eikonal QNMs and the photon ring of the black holes would be broken in the presence of such a general nonminimal matter-geometry coupling. Another possible extension of this work is to investigate the polar perturbations in this model. According to Refs.~\cite{Bhattacharyya:2017tyc,Bhattacharyya:2018qbe,Datta:2019npq}, in the metric variational principle, there is an additional scalar degree of freedom for the $f(R)$ gravity, and this degree of freedom appears as an inhomogeneous source term in the master equation of the polar perturbations. It can be expected that this scalar field degree of freedom will also appear in the gEMSG and in the theory given by Eq.~\eqref{actiongeneralmost}. One can then investigate how the master equations and the polar QNMs are affected in the presence of this additional degree of freedom. Furthermore, it would be interesting to consider the theories constructed upon the Palatini variational principle and study the black hole perturbations in these theories. We will leave these issues for future works.

\acknowledgments

CYC and PC are supported by Ministry of Science and Technology (MOST), Taiwan, through No. 107-2119-M-002-005, Leung Center for Cosmology and Particle Astrophysics (LeCosPA) of National Taiwan University, and Taiwan National Center for Theoretical Sciences (NCTS). CYC is also supported by MOST, Taiwan, through No. 108-2811-M-002-682. PC is in addition supported by U.S. Department of Energy under Contract No. DE-AC03-76SF00515.

\end{document}